\documentclass[%
 reprint,
showpacs,preprintnumbers,
 amsmath,amssymb,
 aps,
]{revtex4-1}

\usepackage{graphicx}
\usepackage{dcolumn}
\usepackage{bm}

\begin{document}


\title{Nucleation pathway of core-shell composite nucleus in size and composition space and in component space}

\author{Masao Iwamatsu}
\email{iwamatsu@ph.ns.tcu.ac.jp}
\affiliation{%
 Department of Physics, Faculty of Liberal Arts and Sciences, Tokyo City University, Setagaya-ku, Tokyo 158-8557, JAPAN
}%




\date{\today}

\begin{abstract}
The kinetics of nucleation of a core-shell composite nucleus that consists of a core of stable final phase surrounded by a wetting layer of intermediate metastable phase is studied using the kinetic theory of binary nucleation not only in the size and composition space but also in the component space. The steady-state solution of the Fokker-Planck equation is considered.   Various formulas for the critical nucleus at the saddle point as well as for the post-critical nucleus are derived.  The kinetics of nucleation at the saddle point is more appropriately characterized in the size and composition space, while the kinetics of the post-critical nucleus is more appropriately described in the component space.  Although both the free-energy landscape and the reaction rates play decisive role to determine the kinetics of nucleation at the saddle point,  the details of the free energy landscape are irrelevant to the kinetics of the post critical nucleus.   
\end{abstract}

\maketitle

\section{\label{sec:sce1}Introduction}
Nucleation is a very basic phenomena which plays a vital role in various material processing applications ranging from steel production to food and beverage industries~\cite{Kelton2010}.  Recently, researchers have focused on the nucleation of complex materials~\cite{Kelton2010}.  The nucleation of such complex materials can also be complex and often involves intermediate metastable phases~\cite{Ostwald1897,Vekilov2012, Erdemir2009, Gebauer2008,Chung2009}, which appears during the course of nucleation and growth as predicted from the Ostwald's step rule~\cite{Ostwald1897}.  Then, the critical nucleus often has a core-shell structure with stable final phase surrounded by an intermediate metastable phase.

Such a core-shell nucleus appears in various circumstances.  It is well known, for example, the nucleation of protein crystal proceeds through the core-shell type nucleus with the final stable crystal surrounded by the metastable dense solution~\cite{Vekilov2012}.  The model calculation~\cite{tenWolde1997}  using the Monte Carlo simulation with a simplified intermolecular interaction revealed that the critical nucleus at the saddle point corresponds indeed to the core-shell structure.  Subsequent numerical simulation using the Lennard-Jones system~\cite{Meel2008} and a model calculation using the capillarity approximation~\cite{Iwamatsu2011} confirmed the core-shell structure of critical nucleus which corresponds to the saddle point of the free-energy landscape.  A similar composite nucleus is predicted even for a simple metal like aluminum~\cite{Desgranges2007} using molecular dynamics simulation.  Experimental evidence of such a core-shell structure of critical nucleus is observed not only in protein crystallization~\cite{Vekilov2012} but also in colloidal crystallization~\cite{Zhang2007,Savage2009}.

A similar composite nucleus with core-shell structure is considered in the problem of delquescence~\cite{Shchekin2008,McGraw2009}, where the condensation of liquid from supersaturated vapor occurs on a soluble core.  This core-shell structure also appears in various problems such as semiconductor nano-crystals~\cite{Fisher2005}, polymer crystallizations~\cite{Liu2011}, and nano-clusters of alloys~\cite{Colin2012}.  It is also used as a model of linked-flux nucleation or partitioning transformation when the interface-limited growth and diffusion of material is coupled~\cite{Russell1968,Kelton2000,Kelton2003,Diao2008}.   In this model, a core of the final stable phase is surrounded by a shell of the dense pre-nucleus environment~\cite{Russell1968}. 

In the previous paper~\cite{Iwamatsu2012b}, we have studied the nucleation flux of composite nucleus with core-shell structure at the saddle point in the free-energy surface using the theory of decay of metastable phase~\cite{Kramers1940,Langer1969,Berezhkovskii2005,Peters2009} and of multicomponent nucleation~\cite{Trinkaus1983,Wilemski1999}.  In this paper, we will supplement our previous study~\cite{Iwamatsu2012b},  and will study the nucleation pathway for the critical as well as the post-critical nucleus with core-shell structure.  We first recapture the Fokker-Planck or the Zeldovich-Frenkel equation~\cite{Kelton2010,Wu1997} for the composite nucleus~\cite{Iwamatsu2012b} from the Master equation (Section \ref{sec:sec2}) by regarding the composite nucleus as a fully phase-separated two-component binary system.  Then the nucleation flux will be characterized more appropriately using the size-composition representation instead of the two-components representation~\cite{Iwamatsu2012b}  (Section \ref{sec:sec3}).  The nucleation pathway of the post-critical composite nucleus will also be studied using the Zeldovich relation~\cite{Zeldovich1943,Frenkel1955}.   Finally, section \ref{sec:sec5} will contain the conclusion of the study, indicating the outcome as well as suggestions on further research in the field.

\section{\label{sec:sec2}Fokker-Planck equation for the composite nucleus}

\begin{figure}[htbp]
\begin{center}
\includegraphics[width=0.60\linewidth]{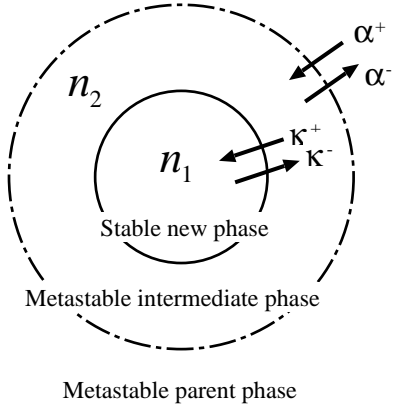}
\end{center}
\caption{
A core-shell critical nucleus model that consists of a stable new phase (number of molecules $n_1$) surrounded by an intermediate metastable phase (number of molecules $n_2$) nucleated in the metastable parent phase~\cite{Iwamatsu2011}.  Transformation rates $\kappa^{+}$ and $\kappa^{-}$ are the reaction rates between the stable new phase and the metastable intermediate phase.  Attachment rates $\alpha^{+}$ and $\alpha^{-}$ are the reaction rate from the metastable parent phase to the intermediate metastable phase.   } 
\label{fig:2x}
\end{figure}

In order to study the nucleation kinetics of the composite nucleus with core-shell structure, the model shown in Fig.~\ref{fig:2x} has been used~\cite{Iwamatsu2011,Iwamatsu2012b}.  The model consists of a core of the stable new phase (number of molecules $n_1$) surrounded by an intermediate metastable phase (number of molecules $n_2$) nucleated in the metastable parent phase.  Kashchiev and coworkers~\cite{Kashchiev1998, Kashchiev2005} have also considered this model as a model of nucleus when there exists an intermediate metastable phase. 

This core-shell nucleus is considered to form by the two-step mechanism: First, the nucleus of the metastable intermediate phase appears within the metastable parent phase.  Next, the core of the stable new phase starts to nucleate inside the nucleus of the metastable intermediate phase to form the core-shell structure shown in Fig.~\ref{fig:2x}.  However,  in contrast to the na\"ive expectation~\cite{Kashchiev2005} the computer simulation~\cite{tenWolde1997}, mean-field calculation~\cite{Iwamatsu2011} as well as the experimental result in colloidal crystallization~\cite{Savage2009} have shown that this two-step mechanism is in fact the single-step nucleation characterized by a single activation energy and, therefore, by a single saddle point.   This saddle point corresponds to the core-shell critical nucleus~\cite{tenWolde1997,Iwamatsu2011,Zhang2007} of the specific size and composition.  Therefore, we will consider the simplest example of the core-shell nucleus of a single-component system.  More complex scenarios will be expected for multi-component system such as the problem of deliquiescence~\cite{Shchekin2008,McGraw2009} and of semiconductor nano-crystals~\cite{Fisher2005}.

Transformation rates $\kappa^{+}$ and $\kappa^{-}$ in Fig.~\ref{fig:2x} are the reaction rates between the stable new phase and the metastable intermediate phase.  Attachment rates $\alpha^{+}$ and $\alpha^{-}$ are the reaction rate from the metastable parent phase to the intermediate metastable phase.  The formation of stable final phase (core) is assume to occur only through the transformation of surrounding metastable phase (shell).

The Master equation for the time-dependence of the concentration of clusters $f\left(n_1,n_2,t\right)$ that consists of $n_1$ molecules of the stable phase in the core and $n_2$ molecules of the intermediate metastable phase  in the surrounding shell is written generally~\cite{Kelton2010,Wu1997, Reiss1950} in the form
\begin{eqnarray}
\frac{\partial f\left(n_1,n_2,t\right)}{\partial t}
&=& \alpha^{+}\left(n_1,n_2-1\right)f\left(n_1,n_2-1,t\right) \nonumber \\
&-&\left[\alpha^{+}\left(n_1,n_2\right)+\alpha^{-}\left(n_1,n_2\right)\right]f\left(n_1,n_2,t\right) \nonumber \\
&+&\alpha^{-}\left(n_1,n_2+1\right)f\left(n_1,n_2+1,t\right) \nonumber \\
&+& \kappa^{+}\left(n_1-1,n_2+1\right)f\left(n_1-1,n_2+1,t\right) \nonumber \\
&-&\left[\kappa^{+}\left(n_1,n_2\right)+\kappa^{-}\left(n_1,n_2\right)\right]f\left(n_1,n_2,t\right) \nonumber \\
&+& \kappa^{-}\left(n_1+1,n_2-1\right)N\left(n_1+1,n_2-1\right), \nonumber \\
\label{eq:1z}
\end{eqnarray}
Using the detailed balance condition 
\begin{eqnarray}
\kappa^{-}\left(n_1+1,n_2-1\right)&=&\kappa^{+}\left(n_1,n_2\right)\frac{f_{\rm eq}\left(n_1,n_2\right)}{f_{\rm eq}\left(n_1+1,n_2-1\right)}, \nonumber \\
\alpha^{-}\left(n_1,n_2+1\right)&=&\alpha^{+}\left(n_1,n_2\right)\frac{f_{\rm eq}\left(n_1,n_2\right)}{f_{\rm eq}\left(n_1,n_2+1\right)},
\label{eq:4z}
\end{eqnarray}
and the usual equilibrium cluster distribution $f_{\rm eq}\left({\bm n}\right)$ given by 
\begin{equation}
f_{\rm eq}\left({\bm n}\right)=f_{0}\exp\left(-\beta G\left({\bm n}\right)\right),
\label{eq:5z}
\end{equation}
where $G\left({\bm n}\right)$ is the work of cluster formation for a cluster with composition ${\bm n}=\left(n_1,n_2\right)$, we obtain
\begin{eqnarray}
&&\frac{\partial f\left(n_1,n_2,t\right)}{\partial t}= \nonumber \\
&-&\kappa^{+}\left(n_1,n_2\right)f_{\rm eq}\left(n_1,n_2\right) \nonumber \\ 
&&\times\left[\frac{f\left(n_1,n_2,t\right)}{f_{\rm eq}\left(n_1,n_2\right)}-\frac{f\left(n_1+1,n_2-1,t\right)}{f_{\rm eq}\left(n_1+1,n_2-1\right)}\right]
\nonumber \\
&+&\kappa^{+}\left(n_1-1,n_2+1\right)f_{\rm eq}\left(n_1-1,n_2+1\right) \nonumber \\
&&\times\left[\frac{f\left(n_1-1,n_2+1,t\right)}{f_{\rm eq}\left(n_1-1,n_2+1\right)} 
-\frac{f\left(n_1,n_2,t\right)}{f_{\rm eq}\left(n_1,n_2\right)}\right] \nonumber \\
&-&\alpha^{+}\left(n_1,n_2\right)f_{\rm eq}\left(n_1,n_2\right) \nonumber \\
&&\times\left[\frac{f\left(n_1,n_2,t\right)}{f_{\rm eq}\left(n_1,n_2\right)} 
-\frac{f\left(n_1,n_2+1,t\right)}{f_{\rm eq}\left(n_1,n_2+1\right)}\right] \nonumber \\
&+&\alpha^{+}\left(n_1,n_2-1\right)f_{\rm eq}\left(n_1,n_2-1\right) \nonumber \\
&&\times\left[\frac{f\left(n_1,n_2-1,t\right)}{f_{\rm eq}\left(n_1,n_2-1\right)} 
-\frac{f\left(n_1,n_2,t\right)}{f_{\rm eq}\left(n_1,n_2\right)}\right].
\label{eq:6z}
\end{eqnarray}
which can be written in the form of continuum equation:
\begin{equation}
\frac{\partial N\left({\bm n}\right)}{\partial t}=-\left(\frac{\partial J_{n_1}}{\partial n_1}+\frac{\partial J_{n_2}}{\partial n_2}\right)=-{\rm div}{\bm J}=-\nabla{\bm J},
\label{eq:10z}
\end{equation}
where the components of the nucleation flux $\bm J$ are given by
\begin{eqnarray}
J_{n_1}&=-&\kappa^{+}f_{\rm eq}\left\{\frac{\partial \Phi}{\partial n_1}
-\frac{\partial \Phi}{\partial n_2}\right\},  \label{eq:11z}
\\
J_{n_2}&=&-\alpha^{+} f_{\rm eq}\left\{\frac{\partial \Phi}{\partial n_2}\right\}  \nonumber \\
&&-\kappa^{+}f_{\rm eq}\left\{-\frac{\partial \Phi}{\partial n_1}
+\frac{\partial \Phi}{\partial n_2}\right\},
\label{eq:12z}
\end{eqnarray}
and simplified notations $\kappa^{+}=\kappa^{+}\left(n_1,n_2\right)$, $\alpha^{+}=\alpha^{+}\left(n_1,n_2\right)$ and $f_{\rm eq}=f_{\rm eq}\left(n_1,n_2\right)$ and $\Phi=\Phi\left(n_1,n_2,t\right)$ are used. 
where
\begin{equation}
\Phi\left(n_1,n_2,t\right)=\frac{f\left(n_1,n_2,t\right)}{f_{\rm eq}\left(n_1,n_2\right)}.
\label{eq:9z}
\end{equation}
Apparently,
\begin{equation}
J_{n_1}+J_{n_2}=-\alpha^{+} f_{\rm eq}\frac{\partial \Phi}{\partial n_2}.
\label{eq:13z}
\end{equation}
Therefore, the net flux coming into this composite nucleus is the incoming flux to the wetting layer of the intermediate phase from surrounding parent phase whose magnitude is determined from the rate constant $\alpha^{+}$, since all materials are suppled from surrounding parent phase.

Equations (\ref{eq:11z}) and (\ref{eq:12z}) can be put in the form of matrix equation using the short-hand notation $\partial_{n_1}=\partial /\partial n_1$ and $\partial_{n_2}=\partial /\partial n_{2}$ as
\begin{equation}
\begin{pmatrix}
J_{n_1} \\
J_{n_2}
\end{pmatrix}
=-f_{\rm eq}
\begin{pmatrix}
\kappa^{+} & -\kappa^{+} \\
-\kappa^{+} & \kappa^{+}+\alpha^{+}
\end{pmatrix}
\begin{pmatrix}
\partial_{n_1} \\
\partial_{n_2} \\
\end{pmatrix}
\Phi,
\label{eq:14z}
\end{equation}
which will be written in short
\begin{equation}
{\bm J}_{\bm n}=-f_{\rm eq}\left({\bm n}\right){\bm R\left({\bm n}\right)}{\bm \nabla}_{\bm n}{\bm \Phi\left({\bm n}\right)},
\label{eq:15z}
\end{equation}
where $\bm J_{\bm n}$, $\bm \Phi$ and ${\bm \nabla}_{\bm n}$ are row vectors, and $\bm R$ is a symmetric square matrix defined through Eq.~(\ref{eq:14z}).  The two flux $J_{n_1}$ and $J_{n_2}$ are linked~\cite{Russell1968} by the non-diagonal rate matrix $\bm R$.  By introducing the unit row vector ${\bm e}_{\bm n}^{\rm T}=\left({\bm e}_{n_1}, {\bm e}_{n_2}\right)$, where the superscript T indicates the transpose vector, the nucleation current vector is written as
\begin{equation}
\vec{J}={\bm e}_{\bm n}^{\rm T}{\bm J}={\bm e}_{n_1}J_{n_1}+{\rm e}_{n_2}J_{n_2}
\label{eq:16z}
\end{equation}
using the unit vectors ${\bm e}_{n_1}$ and ${\bm e}_{n_2}$ along the Cartesian coordinate $\left(n_1,n_2\right)$. 

Since the Fokker-Planck equation given by Eqs.~(\ref{eq:10z})-(\ref{eq:12z}) for the growth of composite nucleus has the same form as that used to study the binary nucleation, we will extend the theory~\cite{Reiss1950,Stauffer1976,Trinkaus1983,Temkin1984,Greer1990,Wu1997,Wilemski1999} developed for the binary nucleation to study the scenario of nucleation and growth of a composite nucleus in the next section.

\section{\label{sec:sec3} Critical nucleus and post critical nucleus}

\subsection{\label{sec:sec3-1} Critical nucleus in the size and composition space}
In this section we extend the theory of nucleation flux of binary nucleation in the size and composition space developed by Fisenko and Wilemski~\cite{Fisenko2004} to the composite nucleus shown in Fig.~\ref{fig:2x} for which the rate matrix $\bm R$ is non-diagonal.  

At this point we introduce a new coordinate, the size $n$ and the composition $x$ defined by
\begin{eqnarray}
n &=& n_1+n_2, \label{eq:1y} \\
x &=& n_2/n.   \label{eq:2y}
\end{eqnarray}
Then the covariant
formulation~\cite{Risken1989,Graham1977,Grabert1980} of the Fokker-Planck equation can be used, and the distribution function is given by
\begin{equation}
\varphi\left(n,x\right)=n f\left(n_1,n_2\right) \label{eq:3y}
\end{equation}
where the factor $n$ comes from the Jacobian $\partial\left(n_1,n_2\right)/\partial\left(n,x\right)=n$.  Then the Fokker-Planck equation (\ref{eq:10z}) in the size-composition space is written as
\begin{equation}
\frac{\partial \varphi\left(n,x\right)}{\partial t}=-\left(\frac{\partial J_{n}}{\partial n}+\frac{\partial J_{x}}{\partial x}\right)=-{\rm div}{\bm J},
\label{eq:4y}
\end{equation}
where the flux components are defined through
\begin{eqnarray}
J_n &=& n\left(J_{n_1}+J_{n_2}\right), \label{eq:5y} \\
J_x &=& \left(1-x\right)J_{n_2}-xJ_{n_1}, \label{eq:6y}
\end{eqnarray}
and Eq.~(\ref{eq:15z}) is transformed into
\begin{equation}
\begin{pmatrix}
J_{n} \\
J_{x}
\end{pmatrix}
=-f_{\rm eq}
\begin{pmatrix}
\tilde{R}_{n,n} & \tilde{R}_{n,x} \\
\tilde{R}_{x,n} & \tilde{R}_{x,x}
\end{pmatrix}
\begin{pmatrix}
\partial_{n} \\
\partial_{x} \\
\end{pmatrix}
\Phi,
\label{eq:7y}
\end{equation}
where the elements of the reaction rate matrix $\tilde{\bm R}$ defined through Eq.~(\ref{eq:7y}) are given explicitly by
\begin{eqnarray}
\tilde{R}_{n,n} &=& n \sum_{i,j}R_{ij}=nR_{\rm tot}, \label{eq:8y} \\
\tilde{R}_{n,x} &=& \tilde{R}_{x,n}=-x\left(R_{11}+R_{21}\right)+\left(1-x\right)\left(R_{12}+R_{22}\right), \label{eq:9y} \\
\tilde{R}_{x,x} &=& \frac{1}{n}\left(x^2 R_{11}+\left(1-x\right)^2 R_{22}-2x\left(1-x\right)R_{12}\right), \label{eq:10y}
\end{eqnarray}
using the elements of the matrix $\bm R$, which will be written in short
\begin{equation}
{\bm J}_{\bm x}=-f_{\rm eq}\left({\bm x}\right){\tilde{\bm R}\left({\bm x}\right)}{\bm \nabla}_{\bm x}{ \Phi\left({\bm x}\right)},
\label{eq:11y}
\end{equation}
where ${\bm J}_{\bm x}$ and ${\bm \nabla}_{\bm x}$ are the column vectors given explicitly in Eq.~(\ref{eq:7y}).  Equation (\ref{eq:11y}) is formally the same as Eq.~(\ref{eq:15z}).  Therefore, various formulas for the steady-state flux in the binary nucleation can be used just by changing the suffices ${\bm n}=(n_1,n_2)$ to ${\bm x}=(n, x)$.  Then, it is apparent from Eqs.~(\ref{eq:8y})-(\ref{eq:10y}) that $\tilde{R}_{n,n}$ plays the r\^ole of Brownian diffusion coefficient in the size space, while $\tilde{R}_{x,x}$ the Brownian diffusion coefficient in the composition space~\cite{Fisenko2004}.  

Since $\tilde{R}_{x,x}\rightarrow 0$ as $n\rightarrow\infty$, the fluctuation of the composition $x$ will be suppressed and $x$ will become constant as the size of the nucleus $n$ increases.  The growth of nucleus is governed mainly by the reaction rate $\tilde{R}_{n,n}$, and only the size of the nucleus increases.  The coupling of the flux in size $J_{n}$ and that in the composition $J_{x}$ disappears when $\tilde{R}_{n,x}=0$, which is attained for the composition
\begin{equation}
x_{k}=\frac{R_{12}+R_{22}}{R_{\rm tot}}.
\label{eq:12y}
\end{equation}
where $R_{\rm tot}$ is defined through Eq.~(\ref{eq:8y}).  The diffusion coefficient in the composition space $\tilde{R}_{x,x}$ given by Eq.~(\ref{eq:10y}) is also minimized for the composition $x_{k}$ given by Eq.~(\ref{eq:12y}).  Therefore, the composition $x_{k}$ is the {\it kinetically optimum composition} of nucleation, which is solely determined from the kinetic factor $\bm R$.
 
Next we will consider the nucleation flux at the saddle point $(n^{*},x^{*})$ in the size and composition space and $(n_{1}^{*},n_{2}^{*})$ in the component space characterized by
\begin{equation}
\left(\frac{\partial G}{\partial n}\right)_{x}=\left(\frac{\partial G}{\partial x}\right)_{n}=0,
\label{eq:16d}
\end{equation}
or
\begin{equation}
\left(\frac{\partial G}{\partial n_1}\right)_{n_2}=\left(\frac{\partial G}{\partial n_2}\right)_{n_1}=0,
\label{eq:16e}
\end{equation}
where these two sets of equations are equivalent.

From Eqs.~(\ref{eq:15z}) and (\ref{eq:11y}), it is obvious that we can use the results of the previous work~\cite{Iwamatsu2012b} simply by replacing the matrix $\bm R$ by $\tilde{\bm R}$.  First, the angle $\omega$ for the direction of the gradient of $\Phi$ in the size and composition space is given by the formula
\begin{equation}
\tan\omega=s\pm\sqrt{s^{2}+r}
\label{eq:13y}
\end{equation}
where
\begin{eqnarray}
s &=& \frac{\tilde{R}_{x,x}\tilde{G}_{x,x}-\tilde{R}_{n,n}\tilde{G}_{n,n}}{2\left(\tilde{R}_{n,x}\tilde{G}_{n,n}+\tilde{R}_{x,x}\tilde{G}_{n,x}\right)},
\label{eq:14y} \\
r &=& \frac{\tilde{R}_{n,n}\tilde{G}_{n,x}+\tilde{R}_{n,x}\tilde{G}_{x,x}}{\tilde{R}_{n,x}\tilde{G}_{n,n}+\tilde{R}_{x,x}\tilde{G}_{n,x}},
\label{eq:15y}
\end{eqnarray}
and
\begin{equation}
\tilde{G}_{n,n}=\frac{\partial^{2}G}{\partial n^{2}},\;\;\tilde{G}_{x,x}=\frac{\partial^{2}G}{\partial x^{2}},\;\;
\tilde{G}_{n,x}=\frac{\partial^{2}G}{\partial n \partial x}
\label{eq:16y}
\end{equation}
are the derivative (Hessian) of the free-energy $G\left(n, x\right)$ in the size and composition space at the saddle point $(n^{*},x^{*})$.  These quantities are related to the derivative in the component space
\begin{equation}
G_{11}=\frac{\partial^{2}G}{\partial n_{1}^{2}},\;\;G_{22}=\frac{\partial^{2}G}{\partial n_{2}^{2}},\;\;
G_{12}=\frac{\partial^{2}G}{\partial n_{1} \partial n_{2}}
\label{eq:16x}
\end{equation}
through
\begin{eqnarray}
\tilde{G}_{n,n}&=&\left(1-x\right)^2 G_{11}+2x\left(1-x\right) G_{12}+x^2 G_{22}, \label{eq:16a} \\
\tilde{G}_{x,x} &=& n^2\left(G_{11}-2 G_{12}+G_{22}\right), \label{eq:16b}\\
\tilde{G}_{n,x} &=& -n\left(1-x\right)G_{11}+n\left(1-2x\right)G_{12}+nx G_{22} \nonumber \\
&&-\frac{\partial G}{\partial n_{1}}+\frac{\partial G}{\partial n_{2}}. \label{eq:16c}
\end{eqnarray}
The last two terms in Eq.~(\ref{eq:16c}) vanish at the critical point from Eq.~(\ref{eq:16e}).

Equations (\ref{eq:13y})-(\ref{eq:15y}) are slightly different in definition from those used by others~\cite{Stauffer1976,Wilemski1999} since we have off-diagonal elements $\tilde{R}_{n,x}=\tilde{R}_{x,n}$.  
Among the sign $\pm$, $+$ sign must be chosen when $\tilde{R}_{n,x}G_{n,n}+\tilde{R}_{x,x}G_{n,x}<0$ otherwise $-$ sign must be chosen.   For a sufficiently large critical nucleus $n^{*}\gg 1$, we have $\tilde{R}_{n,n}\gg\tilde{R}_{n,x}\gg\tilde{R}_{x,x}$.  Also $\tilde{R}_{n,x}=0$ when the composition is the kinetically optimum composition $x^{*}=x_{k}$.  Then, we have $s^2\gg r$ in Eq.~(\ref{eq:13y}), which results in $\omega=0$.  Therefore, the gradient of $\Phi$ (${\bm \nabla}_{\bm x}\Phi$) will be nearly parallel to the size axis $n$ when the size of the critical nucleus $n^{*}$ is large. Therefore, only the size rather than the composition is expected to increase.

Using the result of the previous paper~\cite{Iwamatsu2012b}, we can derive the formula for the angle $\phi$ of the direction of nucleation flux ${\bm J}_{\bm x}$, which is given by
\begin{equation}
\tan\phi=\frac{\tilde{R}_{n,x}+\tilde{R}_{x,x}\tan\omega}{\tilde{R}_{n,n}+\tilde{R}_{n,x}\tan\omega},
\label{eq:17y}
\end{equation}
Again, for a sufficiently large critical nucleus  $n^{*}\gg 1$, we can approximate
\begin{equation}
\tan\phi\sim \frac{\tilde{R}_{n,x}+\tilde{R}_{x,x}\tan\omega}{\tilde{R}_{n,n}}
\label{eq:18y}
\end{equation}
and $\phi\simeq 0$ as $R_{n,n}\gg R_{x,x}$.  Then the nucleation flux  ${\bm J}_{\bm x}$ also becomes parallel to the size axis $n$ as expected.

Finally, the nucleation rate is given by
\begin{equation}
I=f_0 e^{-\beta G^{*}}\sqrt{{\rm det}{\bm \tilde{\bm R}}\frac{\left|\tilde{\lambda}_{1}\right|}{\tilde{\lambda}_2}},
\label{eq:19y}
\end{equation}
where $\tilde{\lambda}_{1}$ and $\tilde{\lambda}_{2}$ are the negative and the positive eigenvalue of the matrix product $\tilde{\bm G}\tilde{\bm R}$ where
\begin{equation}
{\tilde{\bm G}}=
\begin{pmatrix}
\tilde{G}_{n,n} & \tilde{G}_{n,x} \\
\tilde{G}_{x,n} & \tilde{G}_{x,x}
\end{pmatrix},
\label{eq:21y}
\end{equation}
and $\tilde{\bm R}$ is defined through Eq.~(\ref{eq:7y}).  The matrices $\tilde{\bm R}$ and  $\tilde{\bm G}$ must be calculated at the saddle point.   Therefore the last two terms in Eq.~(\ref{eq:16c}) are zero.

It can be shown directly by calculating the eigenvalues $\tilde{\lambda}_{1}$ and $\tilde{\lambda}_{2}$ that they are related to the eigenvalues $\lambda_{1}$ and $\lambda_{2}$ of the matrix products ${\bm G}{\bm R}$ through
\begin{equation}
\tilde{\lambda}_{1,2}=n\lambda_{1,2}.
\label{eq:22y}
\end{equation}
Also, it can be easily shown that ${\rm det}\tilde{\bm R}={\rm det}{\bm R}$.  Therefore, the nucleation rate Eq.~(\ref{eq:19y}) is also given by the original formula
\begin{equation}
I=f_0 e^{-\beta G^{*}}\sqrt{{\rm det}{\bm R}\frac{\left|\lambda_{1}\right|}{\lambda_2}},
\label{eq:23y}
\end{equation}
in the original $(n_1,n_2)$ composition space as expected.

Qualitative assessment of the relative magnitude of the composition fluctuation and the size fluctuation near the saddle point can be possible.  We follow the argument of Fisenko and Wilemski~\cite{Fisenko2004} and write Eq.~(\ref{eq:7y}) as
\begin{eqnarray}
J_{n} &\simeq& -f_{\rm eq}\left(\tilde{R}_{n,n}\frac{\Delta \Phi}{\Delta n}+\tilde{R}_{n,x}\frac{\Delta \Phi}{\Delta x}\right)\nonumber \\
&=& -f_{\rm eq}\tilde{R}_{n,n}\frac{\Delta \Phi}{\Delta n}\left(1+Z\right), \label{eq:22yy} \\
J_{x} &\simeq& -f_{\rm eq}\left(\tilde{R}_{n,x}\frac{\Delta \Phi}{\Delta n}+\tilde{R}_{x,x}\frac{\Delta \Phi}{\Delta x}\right)\nonumber \\
&=& -f_{\rm eq}\tilde{R}_{x,x}\frac{\Delta \Phi}{\Delta x}\left(1+H\right), \label{eq:23yy}
\end{eqnarray}
where
\begin{eqnarray}
Z&=&\frac{\tilde{R}_{n,x}\Delta n}{\tilde{R}_{n,n}\Delta x}, \label{eq:24y} \\
H&=&\frac{\tilde{R}_{n,x}\Delta x}{\tilde{R}_{x,x}\Delta n}, \label{eq:24yy}
\end{eqnarray}
measure the relative magnitude of the cross terms in Eqs.~(\ref{eq:22yy}) and (\ref{eq:23yy}).  Since
\begin{eqnarray}
\Delta n \simeq \sqrt{1/\beta\left|\partial^{2}\Phi\left(n^{*},x^{*}\right)/\partial n^{2}\right|},
\label{eq:25y}\\
\Delta x \simeq \sqrt{1/\beta\left|\partial^{2}\Phi\left(n^{*},x^{*}\right)/\partial x^{2}\right|},
\label{eq:26y}
\end{eqnarray}
are the curvatures of the free-energy landscape at the saddle point, and
\begin{equation}
\tilde{R}_{n,x}=\tilde{R}_{x,n}=R_{\rm tot}\left(x_{k}-x\right),
\label{eq:27y}
\end{equation}
we find
\begin{eqnarray}
Z &\simeq& \frac{\left(x_{k}-x^{*}\right)}{n^{*}}\sqrt{\frac{\partial^{2}\Phi\left(n^{*},x^{*}\right)/\partial x^{2}}{-\partial^{2}\Phi\left(n^{*},x^{*}\right)/\partial n^{2}} }, \label{eq:28y} 
\\
H &\simeq& \frac{n^{*} R_{\rm tot}\left(x_{k}-x^{*}\right)}{\left(1-x^{*}\right)^2 R_{22}+x^{2}R_{11}-2x^{*}\left(1-x^{*}\right)R_{21}} \nonumber \\
&&\times\sqrt{\frac{-\partial^{2}\Phi\left(n^{*},x^{*}\right)/\partial n^{2}}{\partial^{2}\Phi\left(n^{*},x^{*}\right)/\partial x^{2}} }.
\label{eq:29y}
\end{eqnarray}
When $Z\ll 1$ and $H\ll 1$, the cross terms in Eqs.~(\ref{eq:22yy}) can be neglected, and we can study the time scale of the relaxation in size space $\tau_{n}$ by approximating Eqs.~(\ref{eq:4y}) and (\ref{eq:22yy}) by
\begin{equation}
\frac{\partial \varphi}{\partial t} \sim \frac{\Delta\varphi}{\tau_{n}}
\simeq \frac{\partial}{\partial n}f_{\rm eq}\tilde{R}_{n,n}\frac{\partial \Phi}{\partial n}
\sim\frac{\tilde{R}_{n,n}}{n\Delta n^{2}}\Delta\varphi,
\label{eq:30y}
\end{equation}
from Eq.~(\ref{eq:25y}), which yields
\begin{equation}
\tau_{n}\sim\frac{n\Delta n^{2}}{\tilde{R}_{n,n}}=\frac{\Delta n^{2}}{R_{\rm tot}},
\label{eq:31y}
\end{equation}
for the relaxation time in the size space.  Similarly, the time scale of the relaxation in the composition space $\tau_{x}$ is obtained by approximating Eqs.~(\ref{eq:4y}) and (\ref{eq:22yy}) by
\begin{equation}
\frac{\partial \varphi}{\partial t} \sim \frac{\Delta\varphi}{\tau_{x}} 
\simeq \frac{\partial}{\partial x}f_{\rm eq}\tilde{R}_{x,x}\frac{\partial \Phi}{\partial x}
\sim\frac{\tilde{R}_{x,x}}{n\Delta x^{2}}\Delta\varphi,
\label{eq:32y}
\end{equation}
which yields
\begin{equation}
\tau_{x}\sim\frac{n\Delta x^{2}}{\tilde{R}_{x,x}}.
\label{eq:33y}
\end{equation}
Therefore, the parameter
\begin{equation}
W = \frac{\tau_{x}}{\tau_{n}}=\frac{\tilde{R}_{n,n}\Delta x^{2}}{\tilde{R}_{x,x}\Delta n^{2}}=\frac{H}{Z},
\label{eq:34y}
\end{equation}
will determine the relative magnitude of the relaxation times $\tau_{n}$ and $\tau_{x}$.  If  
\begin{eqnarray}
W &\simeq& \frac{\left(n^{*}\right)^{2} R_{\rm tot}}{\left(1-x^{*}\right)^2 R_{22}+\left(x^{*}\right)^{2}R_{11}-2x^{*}\left(1-x^{*}\right)R_{21}} \nonumber \\
&&\times\frac{-\partial^{2}\Phi\left(n^{*},x^{*}\right)/\partial n^{2}}{\partial^{2}\Phi\left(n^{*},x^{*}\right)/\partial x^{2}}.
\label{eq:35y}
\end{eqnarray}
calculated from Eqs.~(\ref{eq:28y}) and (\ref{eq:29y}) satisfies $W\ll 1$, then $\tau_{x}\ll \tau_{n}$, and the Brownian diffusion along the composition axis is the fastest process and that along the size axis is the slowest process. Then the fluctuation of the composition will be suppressed and the nucleation flux will be narrow along the composition axis.  On the other hand, when $W\gg 1$, the nucleation flux will spread along the composition axis even at the saddle point.  In fact, since $W\gg 1$ in the model calculation by Wyslouzil and Wilemski~\cite{Wyslouzil1995} which shows ridge crossing, Fisenko and Wilemski~\cite{Fisenko2004} argued that we can expect some effect related to the ridge crossing when $W\gg 1$.

From Eqs.~(\ref{eq:22yy}) and (\ref{eq:23yy}), we have
\begin{equation}
\frac{J_{x}}{J_{n}}\simeq \frac{\left(1+H\right)\Delta x}{W\left(1+Z\right)\Delta n},
\label{eq:36y}
\end{equation}
and $\left(\Delta x/\Delta n\right)\ll 1$ in usual condition~\cite{Fisenko2004} when the saddle point is located within a narrow deep valley, we may expect $J_{x}/J_{n}\ll 1$.   Then the nucleation flux at the saddle point along the size axis is much larger than that along the composition axis.

Returning to our original problem of composite nucleus for which the reaction matrix $\bm R$ is given by Eq.~(\ref{eq:14z}), we find 
\begin{eqnarray}
\tilde{R}_{n,n} &=& n\alpha^{+}, \label{eq:36y-1} \\
\tilde{R}_{n,x} &=& \left(1-x\right)\alpha^{+}, \label{eq:36y-2} \\
\tilde{R}_{x,x} &=& \frac{1}{n}\left(\kappa^{+}+\left(1-x\right)^{2}\alpha^{+}\right), \label{eq:36y-3}
\end{eqnarray}
and $R_{\rm tot}=\alpha^{+}$, and the kinetically optimum composition is given by $x_{k}=1$ from Eq.~(\ref{eq:12y}).  We need information of the Hessian of the free-energy $G$ to calculate the angle $\omega$ of the gradient of $\Phi$ from Eq.~(\ref{eq:13y}) and the angle $\phi$ of the direction of the nucleation flux from Eq.~(\ref{eq:17y}).  However, it is apparent that $\phi\rightarrow 0$ as $n\rightarrow\infty$ from Eqs.~(\ref{eq:36y-1})-(\ref{eq:36y-3}) and (\ref{eq:18y}).  Nucleation rate is proportional to $I\propto \sqrt{\alpha^{+}\kappa^{+}}$ from Eq.~(\ref{eq:23y})~\cite{Iwamatsu2012b}.  The time scale given by Eqs.~(\ref{eq:31y}) and (\ref{eq:33y}) can be written more explicitly by
\begin{eqnarray}
\tau_{n} &\sim& \frac{\Delta n^{2}}{\alpha^{+}}, \label{eq:36y-4} \\
\tau_{x} &\sim& \frac{n^{2}\Delta x^{2}}{\kappa^{+}+\left(1-x\right)^{2}\alpha^{+}},
\label{eq:26y-5}
\end{eqnarray}
Therefore, the time scale of nucleation $\tau_{n}$ in the size space is determined from the curvature $\Delta n$ of the free energy surface along the size axis and the attachment rate $\alpha^{+}$ from the parent phase to the intermediate metastable phase.  The time scale $\tau_{x}$ in the composition space, on the other hand, is determined from the curvature $n\Delta x$ of the free energy surface along the composition axis and $\alpha^{+}$ as well as the transformation rate $\kappa^{+}$ from the intermediate metastable phase to the final stable phase.  At the kinetically optimum composition $x_{k}=1$, however, the size and the composition is decoupled and $\tau_{x}$ is determined solely from $n\Delta x$ and $\kappa^{+}$.

\begin{figure}[htbp]
\begin{center}
\includegraphics[width=0.8\linewidth]{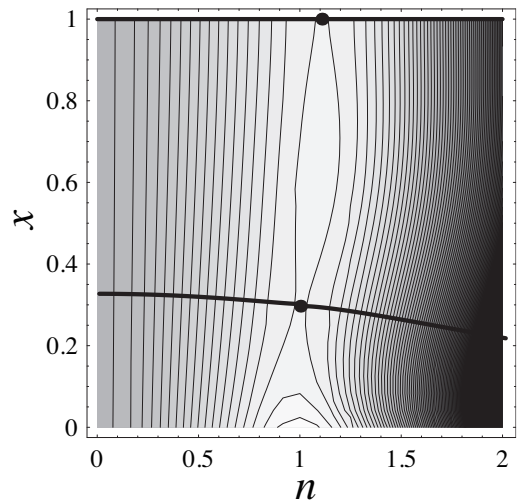}
\caption{
An example of the free-energy surface in the size ($n$) and composition ($x$) space calculated from the model based on the capillarity approximation~\cite{Iwamatsu2011}.  The size $n$ is scaled by the size of the critical nucleus $n^{*}$.  There exist two saddle points indicated by two points.  The one on the $x=1$ axis is the critical nucleus that consists only of the intermediate phase, and the other is the core-shell nucleus that consists of a core of the final stable phase surrounded by a shell of the intermediate phase shown schematically in Fig.~\ref{fig:2x}.  Two heavy lines are two independent nucleation pathways that correspond to the minimum-free-energy paths.   }
\label{fig:3x}
\end{center}
\end{figure}

Figure \ref{fig:3x} shows an example of the free-energy surface calculated from our model~\cite{Iwamatsu2011} in the size $n$ and the composition $x$ space.  The size $n$ is scaled by the typical size of the critical nucleus $n^{*}$.   There are two saddle points of almost the same size with different composition indicated by two points in Fig.~\ref{fig:3x}: the one that corresponds to the critical nucleus of the metastable intermediate phase only ($x=1$) and the other that corresponds to the core-shell nucleus (Fig.~\ref{fig:2x}) that consists of a core of the stable phase surrounded by a shell of the intermediate phase ($x\simeq0.3$).  These two saddle points are not crossed sequentially by a single nucleation pathway.  Rather these are crossed in parallel by two nucleation pathways shown by two heavy lines in Fig.~\ref{fig:3x} independently. These nucleation pathways are the minimum-free-energy path (MFEP)~\cite{Iwamatsu2009} that corresponds to the steepest-descent direction of the free-energy surface.

In our model\cite{Iwamatsu2011}, figure~\ref{fig:3x} seems to suggest $\Delta n/n \sim \Delta x$ since the size $n$ is scaled by the typical size of critical nucleus $n^{*}\sim 10-1000$, but we also need the magnitude of the attachment rate $\alpha^{+}$ and the transformation rate $\kappa^{+}$ to determine the time scale $\tau_n$ and $\tau_x$ of the nucleation quantitatively around these critical points from Eqs.~(\ref{eq:36y-4}) and (\ref{eq:26y-5}).

\subsection{\label{sec:sec3-2} Post-critical nucleus in the component space}

So far, we have considered the critical nucleus at the saddle point.  In order to study the nucleation pathway of not only the nucleation but also the growth of post-critical nucleus, it is more convenient to study Eq.~(\ref{eq:15z}) in the original $(n_1,n_2)$ component space.  This equation is rewritten in the form
\begin{equation}
{\bm J}_{\bm n} = -f_{\rm eq}{\bm R}{\bm \nabla}_{\bm n}{\bm \Phi}= -{\bm R}{\bm \nabla}_{\bm n}f+f \dot{\bm{n}}
\label{eq:37y}
\end{equation}
where the first term in the right-hand side is termed the diffusion flux and the second term the drift flux.  The latter is given explicitly by
\begin{equation}
\dot{\bm{n}}=-\beta{\bm R}{\bm \nabla}_{\bm n}G,
\label{eq:38y}
\end{equation}
which determine the growth of post-critical nucleus though this drift flux is negligible compared to the diffusion flux near the saddle point as $\partial G/\partial n_{i}=0$.  Eq.~(\ref{eq:38y}) is known as the Zeldovich relation~\cite{Zeldovich1943,Frenkel1955}.

Explicitly, Eq.~(\ref{eq:38y}) is written by
\begin{eqnarray}
\dot{n}_{1} &=& -\beta\left(R_{11}\frac{\partial G}{\partial n_{1}}+R_{12}\frac{\partial G}{\partial n_{2}}\right),
\label{eq:39y} \\
\dot{n}_{2} &=& -\beta\left(R_{12}\frac{\partial G}{\partial n_{1}}+R_{2}\frac{\partial G}{\partial n_{2}}\right).
\label{eq:40y} 
\end{eqnarray}
However, as pointed out by Stauffer~\cite{Stauffer1976}, the continuous equation (\ref{eq:15z}) is originally derived from the discrete Master equation (\ref{eq:6z}) of the attachment and the detachment of monomers, therefore we should use the difference equations
\begin{eqnarray}
-\frac{\partial \beta G}{\partial n_{1}} &=& \frac{1}{f_{\rm eq}}\frac{\partial f_{\rm eq}}{\partial n_{1}} \nonumber \\
&\simeq& 1-e^{\beta\left(G\left(n_{1}+1,n_{2}\right)-G\left(n_{1},n_{2}\right)\right)}, \label{eq:41y} \\
-\frac{\partial \beta G}{\partial n_{2}} &=& \frac{1}{f_{\rm eq}}\frac{\partial f_{\rm eq}}{\partial n_{2}} \nonumber \\
&\simeq& 1-e^{\beta\left(G\left(n_{1},n_{2}+1\right)-G\left(n_{1},n_{2}\right)\right)}, 
\label{eq:42y}
\end{eqnarray}
instead of the partial derivatives $\partial G/\partial n_1$ and $\partial G/\partial n_2$.  Then, Eqs.~(\ref{eq:39y}) and (\ref{eq:40y}) are written more appropriately by
\begin{eqnarray}
\dot{n}_{1} &=& R_{11}\left(1-e^{\beta\left(G\left(n_{1}+1,n_{2}\right)-G\left(n_{1},n_{2}\right)\right)}\right) \nonumber \\
&+& R_{12}\left(1-e^{\beta\left(G\left(n_{1},n_{2}+1\right)-G\left(n_{1},n_{2}\right)\right)}\right), 
\label{eq:43y} \\
\dot{n}_{2} &=& R_{12}\left(1-e^{\beta\left(G\left(n_{1}+1,n_{2}\right)-G\left(n_{1},n_{2}\right)\right)}\right) \nonumber \\
&+& R_{22}\left(1-e^{\beta\left(G\left(n_{1},n_{2}+1\right)-G\left(n_{1},n_{2}\right)\right)}\right). 
\label{eq:44y}
\end{eqnarray}

For sufficiently large clusters ($n_{1}\rightarrow \infty, n_{2}\rightarrow \infty$), we can neglect the surface tension of the spherical nucleus and the free energy of the nucleus is approximately given by~\cite{Stauffer1976} $G\left(n_{1}, n_{2}\right)\simeq -\Delta\mu_{1}n_{1}-\Delta\mu_{2}n_{2}$, where $\Delta\mu_{1}>0$ and $\Delta\mu_{2}>0$ are the chemical potential of the stable phase and the metastable intermediate phase relative to the metastable parent phase.  Then, Eqs.~(\ref{eq:43y}) and (\ref{eq:44y}) are approximately given by
\begin{eqnarray}
\dot{n}_{1} &=& R_{11}\left(1-e^{-\beta\Delta\mu_{1}}\right)+R_{12}\left(1-e^{-\beta\Delta\mu_{2}}\right), 
\label{eq:45y} \\
\dot{n}_{2} &=& R_{12}\left(1-e^{-\beta\Delta\mu_{1}}\right)+R_{22}\left(1-e^{-\beta\Delta\mu_{2}}\right). 
\label{eq:46y}
\end{eqnarray}
Therefore, the angle $\phi$ of the direction of the nucleation flux is given by
\begin{eqnarray}
\tan\phi &=& \frac{\dot{n}_{2}}{\dot{n}_{1}} \nonumber \\
&=& \frac{R_{12}\left(1-e^{-\beta\Delta\mu_{1}}\right)+R_{22}\left(1-e^{-\beta\Delta\mu_{2}}\right)}{R_{11}\left(1-e^{-\beta\Delta\mu_{1}}\right)+R_{12}\left(1-e^{-\beta\Delta\mu_{2}}\right)}.
\label{eq:47y}
\end{eqnarray}
Since the number of molecule changes linearly in time $t$ as $n_{1}=\dot{n}_{1}t$ and $n_{2}=\dot{n}_{2}t$, the final composition
\begin{equation}
x_{2}=\frac{n_{2}}{n_{1}+n_{2}}
\label{eq:48y}
\end{equation}
satisfies
\begin{equation}
\frac{x_{2}}{1-x_{2}}=\frac{n_{2}}{n_{1}}=\frac{\dot{n}_{2}}{\dot{n}_{1}}=\tan\phi
\label{eq:49y}
\end{equation}
given by Eq.~(\ref{eq:47y}).  For large supersaturations $\Delta\mu_{1}\rightarrow \infty$ and $\Delta\mu_{2}\rightarrow \infty$, we find
\begin{equation}
\frac{x_{2}}{1-x_{2}}=\tan\phi\rightarrow \frac{R_{12}+R_{22}}{R_{12}+R_{11}}=\frac{x_{k}}{1-x_{k}}
\label{eq:50y}
\end{equation}
Therefore, the final composition becomes the kinetically optimal composition $x_{k}$ given by Eq.~(\ref{eq:12y}).

Returning to our original problem with the reaction matrix $\bm R$ given by Eq.~(\ref{eq:14z}), we find $R_{12}+R_{22}=\alpha^{+}$ and $R_{12}+R_{11}=0$.  Equation (\ref{eq:50y}) predicts that the final composition is $x_{2}=x_{k}=1$.  Then, the post-critical nucleus consists only of the metastable intermediate phase and cannot reach the final stable phase with $x=0$.  

These seemingly unphysical results are due to the assumption that both the stable phase and the metastable intermediate phase are comparably stable ($\Delta\mu_{1}\rightarrow \infty, \Delta\mu_{2}\rightarrow \infty$).  In fact, we should have $\Delta\mu_1> \Delta\mu_2\sim 0$. Using the explicit form $\bm R$ in Eq.~(\ref{eq:14z}), equations (\ref{eq:45y}) and (\ref{eq:46y}) are given by
\begin{eqnarray}
\dot{n}_{1} &=& \kappa^{+}\left(e^{-\beta\Delta\mu_2}-e^{-\beta\Delta\mu_1}\right), \label{eq:51y}
\\
\dot{n}_{2} &=& -\kappa^{+}\left(e^{-\beta\Delta\mu_2}-e^{-\beta\Delta\mu_1}\right)+\alpha^{+}\left(1-e^{-\beta\Delta\mu_2}\right), \label{eq:52y}
\end{eqnarray}
Therefore, the number of molecules $n_{1}$ of the stable phase will increase as $\dot{n}_{1}>0$, while $n_2$ of the intermediate metastable phase may increase or decrease depending not only on the chemical potentials $\Delta\mu_1$ and $\Delta\mu_2$ but also on the reaction rates $\alpha^{+}$ and $\kappa^{+}$  In the late stage of the post-critical nucleus when the depletion of monomer starts to occur, the incoming flux will stop and $\alpha^{+}\simeq 0$, and the second term of Eq.~(\ref{eq:52y}) can be neglected.  Then the number of molecules $n_{2}$ of the metastable intermediate phase will decrease to compensate the increase of the number of molecules $n_1$ of the final stable phase since $\dot{n}_{2}=-\dot{n}_{1}<0$.  Finally the supercritical nucleus will composed only of the molecules of the stable phase $n_1$, and the final stable phase with $x=0$ will be reached.   

In contrast to two minimum-free-energy paths indicated by two heavy curves in Fig.~\ref{fig:3x}, the two real nucleation paths should gradually turn toward $x=0$ axis after they overcome saddle points indicated by two points.  They should merge into the $x=0$ axis as the size $n$ increases according to Eqs.~(\ref{eq:51y}) and ({\ref{eq:52y}) since $x=0$ ($n_2=0$) in the stable bulk phase. This evolution of the supercritical nucleus will not involve saddle points and, therefore, must be a barrier-less gradual change.  The time scale of evolution is determined mainly by the reaction rate $\kappa^{+}$ and $\alpha^{+}$ from Eqs.~(\ref{eq:51y}) and ({\ref{eq:52y}).  Although both the free-energy landscape and the reaction rates play decisive role to determine the kinetics of nucleation at the saddle point,  the details of the free energy landscape are irrelevant to the kinetics of the post critical nucleus. 

Of course, if the metastable phase is as stable as the final stable phase ($\Delta\mu_1\gtrsim \Delta\mu_2$), prediction of Eq.~(\ref{eq:50y}) suggests that the nucleus made of metastable phase (upper nucleation route with $x=1$ in Fig.~\ref{fig:3x}) could have a long life time since the kinetically optimum composition is $x_k=1$.  Then the metastable intermediate phase would be macroscopically observable.  The observability of the metastable phase around the core stable phase as the core-shell structure  (lower nucleation route in Fig.~\ref{fig:3x}) also depends sensitively on the difference of the chemical potential $\Delta\mu_1$ and $\Delta\mu_2$ from Eq.~(\ref{eq:52y}).

\section{\label{sec:sec5}Conclusion}

In this paper, we have studied the nucleation pathway of the critical and the post critical composite nucleus with core-shell structure not only in the size and composition space but also in the component space.  By extending the results of the previous paper~\cite{Iwamatsu2012b}, we could study the critical nucleus at the saddle point in the size and composition space.  Our results suggested that the critical nucleus can be more appropriately characterized in the size and composition space.  However, the kinetics of post-critical nucleus can be studied more easily in the original component space.

Recent theoretical ~\cite{Iwamatsu2011,tenWolde1997,Meel2008,Desgranges2007} as well as experimental~\cite{Savage2009, Liu2011} results suggest that the appearance of the composite core-shell nucleus does not necessarily mean two successive activations by crossing two saddle points sequentially~\cite{Kashchiev1998, Kashchiev2005, Zhang2007,Vekilov2012}.  Rather, the word two-step simply means that the nucleation pathway takes a roundabout course on the free energy landscape via the single saddle point which corresponds to the composite nucleus with core-shell structure.  Our result in this paper will be useful to understand the nucleation pathway of such a two-step nucleation not only at the saddle point but also at the late stage of growth after crossing the saddle point.

It must be noted, however, our analysis is completely confined to the steady-state process.  The transient properties can only be studied numerically by solving coupled Master Equations~\cite{Kelton2000,Diao2008,Wyslouzil1996}.  In addition, our analysis is conducted under the assumption that the nucleation flux goes through the saddle point.  Saddle point avoidance~\cite{Trinkaus1983,Zitserman1990,Berezhkovskii1995,Wyslouzil1995} will be important if the anisotropy of the reaction matrix $\bm R$ is large or the ridge between saddle point is low, which can occur at high temperatures or near the spinodal point.  In such a case, the ridge-crossing rather than the saddle-crossing may occur.  Then the nucleation flux will spread over the whole phase space and the picture used in this study may break down.  Also the growth of composite nucleus will be expected to be more complex.  Finally, it should be noted that the recent progress to extract the reaction matrix $\bm R$ from simulation data~\cite{Ma2006, Peters2009} makes it possible to apply our formulation directly to a more realistic situation.

\begin{acknowledgments}
This work was supported by the Grant-in-Aid for Scientific Research [Contract No.(C)22540422] from Japan Society for the Promotion of Science (JSPS) and the MEXT supported program for the Strategic Research Foundation at Private Universities, 2009-2013. 
\end{acknowledgments}


\end{document}